\documentclass[14pt]{article}

\usepackage{arxiv}
\usepackage[utf8]{inputenc} 
\usepackage[T1]{fontenc}    
\usepackage{graphicx}
\usepackage{tabularx}

\usepackage{xcolor}

\usepackage{mathptmx}      
\usepackage{flushend}
\usepackage{amsmath}	
\usepackage{amssymb}	
\usepackage{threeparttable}
\usepackage{newtxtext,newtxmath}
\usepackage{booktabs,caption}
\usepackage{multirow}
\usepackage{float}
\usepackage{rotating}
\usepackage{etoolbox}
\usepackage[colorlinks,citecolor=blue,urlcolor=blue,linkcolor=red]{hyperref}

\title{Dark Matter Effects on Compact Binary Stars}

\author{
  Ebrahim Hassani \\
  Department of Physics, Faculty of Sciences, University of Birjand, Birjand, Iran \\
  \texttt{\url{ebrahim.hassani@birjand.ac.ir, eb.hassani7@gmail.com}} \\
  \And
  Reza Pazhouhesh \\
  Department of Physics, Faculty of Sciences, University of Birjand, Birjand, Iran \\
   \texttt{\url{rpazhouhesh@birjand.ac.ir}} \\
}

\begin{document}
	
\maketitle


\begin{abstract}
	\begin{description}
		\item[Abstract]
As compact binary star systems move inside the halo of the galaxies, they interact with dark matter particles. The interaction between dark matter particles and baryonic matter causes dark matter particles to lose some part of their kinetic energy. Once dark matter particles lose part of their kinetic energy, they are gravitationally bound to stars, and stars start to absorb dark matter particles from the halo. The absorption of dark matter particles inside compact binary systems increases the mass of the binary components, and then the total mass of the binary systems increases as well. According to Kepler's third law, increased mass through this way can affect other physical parameters (e.g., semi-major axes and orbital periods) of these systems too. We estimate the period change of some known compact binary systems due to the accretion of dark matter particles into them. We investigate the effects of different dark matter particle candidates with masses in the range $\simeq 10^{-9} - 10^{5} \: (GeV.c^{-2})$ and dark matter density as high as the dark matter density near the Galactic central regions. We find that the estimated change of period due to the accretion of dark matter particles into compact binary systems can be as large as the measured values for these systems.
	\end{description}
\end{abstract}

\section{Introduction} \label{Introduction}
Rotation curves of galaxies reveal the non-uniform distribution of dark matter (DM) inside them \cite{Sofue2001}. From this, one can infer that all astronomical objects inside galaxies are immersed inside DM. Therefore, one can argue that the presence of DM affects the physics of every astronomical object, including stars, binary star systems, and stellar clusters inside galaxies. \\
The interaction of DM particles with baryonic matter through weak interaction can affect the internal structure and the evolutionary courses of stars. According to the known properties of DM particles, DM particles can affect the physics of stars mostly in two ways \cite{2009MNRAS.394...82S,Turck-Chieze2012}: 1) DM particles can transfer energy (whether DM particles annihilate or not) between different layers of stars. 2). If DM particles annihilate inside stars, they can act as a new source of energy (beside the energy that comes from baryonic matter energy production cycles). In other words, considering the effects of DM on the physics of stars causes them to follow different evolutionary paths compared to the standard stellar evolutionary models. \\
In 1978, Steigman proposed DM effects as a solution to the solar neutrino problem \cite{1978AJ.....83.1050S} (though, after discovering the neutrino oscillations by the Super-Kamiokande experiment in 1998 \cite{PhysRevLett.81.1562} this problem has been considered as a solved problem by the neutrino oscillations assumption, instead of the DM assumption \cite{Fisher1999}). Simulation of dwarf galaxies also supports the interaction between DM particles and individual stars. In evolved dwarf galaxies, DM halo around dwarf galaxies heats up by the stars inside them. Then, the more the dwarf galaxy evolves, the more the DM halo heats up by the stars inside them \cite{Read_2018fxs}. In addition, the effects of DM on stars can be used to solve the paradox of youth problem for stars that are located near the central regions of our galaxy \cite{Hassani_2020zvz}. In addition to normal stars, DM effects on other celestial bodies like the moon \cite{2020PhRvD.102b3024C,2020PhLB..80435403G}, planets \cite{Leane2021,2012JCAP...07..046H}, neutron stars (NS) \cite{Rezaei_2017,REZAEI20181,2018JCAP...09..018B, 2018JHEP...11..096K,2018ApJ...863..157C,2013PhRvD..87l3507B,Raj2018,Joglekar2020,Baryakhtar2017}, white dwarf stars (WD) \cite{2019JCAP...08..018D,2018PhRvD..98f3002C, 2018PhRvD..98k5027G,2016MNRAS.459..695A}, black holes \cite{2012PhRvD..85b3519M,2009JCAP...08..024U,Belotsky2014} and binary star systems \cite{Hassani2020b} have been investigated in the literature. \\
As time passes, stars absorb and gather DM particles from the halo of galaxies. As a result, the mass of the stars increases. According to the definition, the capture rate (CR) of DM particles by a massive spherical body (like Earth, Sun, neutron stars, etc.) is the number of DM particles that are gravitationally bound to that body per unit time \cite{Gould_1987ir}. For the first time, Press and Spergel estimated the CR relation of weakly interacting massive particles (WIMP) particles that are accreting into the sun \cite{1985ApJ...296..679P}. In the next step, Gould obtained a general relation for the capture rate of DM particles by other round bodies like planets and stars \cite{Gould_1987ir}. Kouvaris used Press and Spergel relation to derive CR relation for NSs \cite{2008PhRvD..77b3006K}. Hurst et al. used Gould relation to obtain CR relation for WDs \cite{2015PhRvD..91j3514H}.\\
As a compact binary system moves inside a DM halo, it gets affected by the dynamical friction that is imposed by the DM halo. Induced dynamical friction by the DM halo can alter the physical parameters (e.g. orbital period and semi-major axis) of these systems. Dynamical friction that is imposed in this way can be used to constrain parameters of different DM models \cite{GOMEZ2019100343}. In a series of studies, people have tried to estimate the change of period of compact binary systems due to the dynamical friction that is imposed by DM \cite{PhysRevD.96.063001, Penarrubia2016,Blas2020, GOMEZ2019100343, Blas2017,2015PhRvD..92l3530P, Armaleo2020, CAPUTO20181, Yoo2004}. As an example, Gabriel and Rueda compared the observed period change of some known compact binary systems with the estimated period change due to dynamical friction and also the estimated period change due to the gravitational waves emissions from these systems \cite{PhysRevD.96.063001}. One of their results is that, in some regions of parameter space of DM, the orbital decay due to dynamical friction could be comparable to the orbital decay due to the gravitational wave emission. \\
In addition to "dynamical friction" and "gravitational wave emission," the accretion of DM particles inside compact binary systems can be the source of period change (and other physical parameters, like semi-major axis) in these systems. As a compact binary system moves inside the DM halo, each star of the system can capture DM particles from the DM halo. The captured DM particles increase the total mass of the system as well. Then, according to Kepler's third law ( i.e., $ P^{2} \varpropto a^{3}/M $), the increased total mass of the binary systems through this way, alters the semi-major axis and orbital period of these systems. \\
In Sec. (\ref{Sec_Results_and_discussion}), we estimate the period change of some known compact binary systems due to the accretion of DM particles inside them and then compare the results with the period change of these systems due to dynamical friction and gravitational waves emission (see table (\ref{table_CR_in_known_binary_systems}) for a summary of main results). The rest of the paper is structured as follows,  in Sec. (\ref{Sec: Capture_Rate_Relations}), we present and discuss CR relations for white dwarfs and neutron stars. In Sec. (\ref{Sec_CR_By_Compact_Binary_Systems}), a relation for the CR by compact binary systems is obtained.  In Sec. (\ref{Sec: Dark Matter Density Profile}), we explain DM density distribution inside galaxies. Finally, Sec. (\ref{Sec_Results_and_discussion}) is devoted to the results and conclusions. \\
\section{Capture Rate by Compact Stars} \label{Sec: Capture_Rate_Relations}
CR of DM particles by hydrogen atoms is different from the CR relation for elements heavier than hydrogen atoms. Since in hydrogen atoms, the role of the spin of the outermost electron is not insignificant, it should be taken into account in scattering cross section relations. CR by Hydrogen atoms is given by \cite{Hassani2020b}
\begin{multline} \label{Eq: CR_by_hydrogen}
C_{\chi ,H} = \left [ 4\sqrt{6\pi } \frac{\rho_{\chi}}{m_{\chi}} \frac{1}{\overline{v}_{\chi }v_{\ast}} exp(-\frac{3v^{2}_{\ast}}{2\overline{v}^{2}_{\chi}}) \right ] \times
\left [ \sigma_{\chi,SI} + \sigma_{\chi,SD} \right ] \times
\left [ \int_{0}^{R_{\ast}} n_{H}(r) r^{2} dr \right ] \times \\
\left [ \int_{0}^{\infty } exp(-\frac{3u^{2}}{2\overline{v}^{2}_{\chi}}) sinh(\frac{3uv_{\ast}}{\overline{v}^{2}_{\chi}}) (v_{e}^{2}-\frac{\mu_{-,H}^{2}}{\mu_{H}}u^{2}) \theta (v_{e}^{2}-\frac{\mu_{-,H}^{2}}{\mu_{H}}u^{2}) du \right ],
\end{multline} 
and for elements heavier than Hydrogen it has the following form \cite{Hassani2020b}
\begin{multline} \label{Eq: CR_by_hevier}
C_{\chi ,i} = \left [ 8\sqrt{6\pi } \frac{\rho_{\chi}}{m_{\chi}^{2}} \frac{E_{0}}{\overline{v}_{\chi }v_{\ast}} \frac{\mu^{2}_{+,i}}{\mu_{i}} exp(-\frac{3v^{2}_{\ast}}{2\overline{v}^{2}_{\chi}}) \right ] \times
\left [ \sigma_{\chi,SI} A_{i}^{2} (\frac{m_{\chi}m_{n,i}}{m_{\chi}+m_{n,i}})^{2}(\frac{m_{\chi}+m_{p}}{m_{\chi}m_{p}})^{2} \right ] \times
\left [ \int_{0}^{R_{\ast}} n_{i}(r) r^{2} dr \right ] \times \\
[ \int_{0}^{\infty } exp(-\frac{3u^{2}}{2\overline{v}^{2}_{\chi}}) sinh(\frac{3uv_{\ast}}{\overline{v}^{2}_{\chi}}) \: \times
  \left \{ exp(-\frac{m_{\chi}u^{2}}{2E_{0}}) - exp(-\frac{m_{\chi}u^{2}}{2E_{0}}\frac{\mu_{i}}{\mu^{2}_{+,i}}) exp(-\frac{m_{\chi}v_{e}^{2}}{2E_{0}}\frac{\mu_{i}}{\mu^{2}_{-,i}} (1-\frac{\mu_{i}}{\mu^{2}_{+,i}})) \right \}   du  ].
\end{multline}
In Eqs. (\ref{Eq: CR_by_hydrogen}) and (\ref{Eq: CR_by_hevier}), $ \rho_{\chi} $ is the DM density that surrounds the star, $ m_{\chi} $ is the mass of DM particles, $ \overline{v}_{\chi } $ is the dispersion velocity of DM particles, $ v_{\ast} $ is the speed of star relative to the DM halo, $ \sigma_{\chi,SI} $ is the spin-independent scattering cross-section, $ \sigma_{\chi,SD} $ is the spin-dependent scattering cross-section, $ n_{H} $ is the number density of Hydrogen atoms in different locations of stars, $ n_{i} $ is the number density of heavier (i.e. heavier than Hydrogen atoms) elements in different locations of stars, $ r $ is the distance from the center of the star, $ m_{p} $ is the mass of protons, $ m_{i} $ is the nuclear mass of the element i, $ A_{i} $ is the atomic number of element i, $ E_{0} = 3 \hslash ^{2} / (2 m_{n,i}(0.91 m_{n,i}^{1/3}+0.3)^{2}) $ is the characteristic coherence energy (for more details see Ref. \cite{Gould_1987ir}) and $ \mu_{i} $, $\mu_{+,i}$ and $\mu_{-,i}$ are defined in the form:$ \mu_{i} = m_{\chi}/m_{n,i} $ and $ \mu_{\mp,i} \equiv (\mu_{i}\mp1 )/{2} $. \\
As mentioned above, in Eqs. (\ref{Eq: CR_by_hydrogen}) and (\ref{Eq: CR_by_hevier}), quantities $ \sigma_{\chi,SI} $ and $ \sigma_{\chi,SD} $ represents  spin-independent and spin-dependent scattering cross sections respectively. Scattering cross section is a measure of the rate (or probability) of the scattering of DM particles from baryonic matter particles. In practice, scattering cross section depends on many physical parameters, like the structure of the particles (atoms, molecules, photons, etc.) under interaction, whether the interaction is elastic or inelastic, the relative velocity of the particles under the interaction, charge, spin, mass of the particles under interaction, and etc. In addition, due to the lack of knowledge about the nature of DM particles, the exact definition of scattering cross section for every DM particle candidate is highly model-dependent. For example, the definition of scattering cross section for SuperWIMP (one of DM candidates) differs from model to model. To surmount these difficulties, we select maximal DM particle-nucleon scattering cross sections allowed by the direct DM detection experiments. Similar studies used the same reasoning to chose numerical values for DM particle-nucleon scattering cross-section \cite{Lopes2011,2019ApJ...879...50L,Zentner2011b,2009ApJ...705..135C}. For DM candidates that are used in this study (i.e. Sterile Neutrinos, SuperWIMP, WIMP), we choose scattering cross-sections to be: $ \sigma_{Sterile Neutrino,SI} = 10^{-42} cm^{2}$ for Sterile Neutrinos \cite{Formaggio2012} and $ \sigma_{SuperWIMP,SI} = \sigma_{WIMP,SI} = 10^{-44} cm^{2}$ for SuperWIMP and WIMP particles, which are the maximum magnitudes that are determined through the experimental DM detection experiments \cite{2008PhRvL.100b1303A, 2008PhRvL.101i1301A}. \\
In evolved stars, like white dwarfs and neutron stars, stars consume almost all of their initial Hydrogen content. Thus, in compact stars, the contribution of Hydrogen atoms in capturing DM particles is insignificant compared to the heavier elements. Therefore, we only use Eq. (\ref{Eq: CR_by_hevier}) to calculate the CR of DM particles by white dwarfs and neutron stars. In addition, since in elements heavier than Hydrogen, the contribution of the spin of the atoms in the scattering process is not significant (in comparison with the spin part of the scattering cross-section), we only consider the spin-independent part of the scattering cross section (i.e. $ \sigma_{\chi,SI} $) in our calculations. \\
Eq. (\ref{Eq: CR_by_hevier}) consists of four different brackets. The first three brackets can be calculated analytically. We calculate the fourth bracket using numerical integration. In third bracket, the amount of number density is supposed to be a constant value for both white dwarfs and neutron stars. The average number density of a typical white dwarf (e.g. Sirius B WD star with mass $M_{WD} = M_{sirius \: B} = 1.018 \: M_{\odot}$ \cite{Bond2017} and radius $ R_{WD} = R_{sirius \: B} = 0.0084 \: R_{\odot} $ \cite{Holberg1998}) is approximately
\begin{equation} \label{Eq: Number_Density_White_Dwarf}
\bar{n}_{WD} = \dfrac{N}{V} = \dfrac{(\dfrac{M_{WD}}{m_{n}})}{(4/3) \pi R^{3}_{WD}} \simeq  1.4 \times 10^{36} \: (kg.m^{-3}).
\end{equation}
In Eq. (\ref{Eq: Number_Density_White_Dwarf}) $m_{n}$ is the mass of neutron particles and is considered to be $ m_{n} = 1.67 \times 10^{-27} (kg) $. Similar to Eq. (\ref{Eq: Number_Density_White_Dwarf}), the number density of a typical neutron star (i.e. with $M_{NS} = 1.44 \: M_{\odot}$ and $ R_{NS} = 10 \: (km) $ values) can be estimated as
\begin{equation} \label{Eq: Number_Density_Neutron_Stars}
\bar{n}_{NS} = \dfrac{N}{V} = \dfrac{(\dfrac{M_{NS}}{m_{n}})}{(4/3) \pi R^{3}_{NS}} \simeq  4.1 \times 10^{44} \: (kg.m^{-3}).
\end{equation}
\section{Capture rate by compact binary systems} \label{Sec_CR_By_Compact_Binary_Systems}
According to Kepler’s third law, the square of the period in binary systems is proportional to the inverse of the mass of the system and also is proportional to the cube of the semi-major axis of the orbit \cite{hilditch_2001}
\begin{equation} \label{Eq_Keplers_third_law}
P^{2} = (\dfrac{4 \pi^{2}}{GM}) \: a^{3}, 
\end{equation}
where $M=M_{1}+M_{2}$ . If we suppose that all binary systems inside the galaxies are immersed inside the DM halos, then one can assume that both components of the compact binary systems absorb and accrete DM particles. Therefore, as time passes, the total mass of the binary systems will increase. At this point, we estimate the effects of the increased mass on other parameters of the compact binary systems. Taking differential from both sides of the Eq. (\ref{Eq_Keplers_third_law}) (and assuming that there is not a net external force on the system and assuming the absence of other physical effects (e.g. dynamical friction, gravitational waves emission, mass transfer between the components of the binary and, etc) that can affect the parameters of the binary systems) and after some simplifications, we have 
\begin{equation} \label{Eq_CR_by_binary_systems}
\frac{\dot{P}}{P} = \frac{3}{2} \frac{\dot{a}}{a} - \frac{1}{2} \frac{\dot{M}}{M} = \frac{3}{2} \frac{\dot{a}}{a} - \frac{1}{2} \frac{\dot{M_{1}}+\dot{M_{2}}}{(M_{1} + M_{2})},
\end{equation}
where $\dot{P} = dp/dt$, $\dot{a} = da/dt$, $\dot{M} = dM/dt$, $\dot{M_{1}} = dM_{1}/dt$ and $\dot{M_{2}} = dM_{2}/dt$. $\dot{M_{1}}$ and $\dot{M_{2}}$ signifies the mass variation of the binary components due to the accretion of DM particles inside them. To calculate $\dot{M_{1}}$ and $\dot{M_{2}}$ in Eq. (\ref{Eq_CR_by_binary_systems}) we need to multiply Eq. (\ref{Eq: CR_by_hevier}) by the mass of the DM particles $m_{\chi}$ (i.e. $\dot{M_{1}} = CR \times m_{\chi}$ where $M_{1}$ is the mass of the primary compact star). \\
\begin{figure*}
\centering
\includegraphics[width=1.0 \columnwidth]{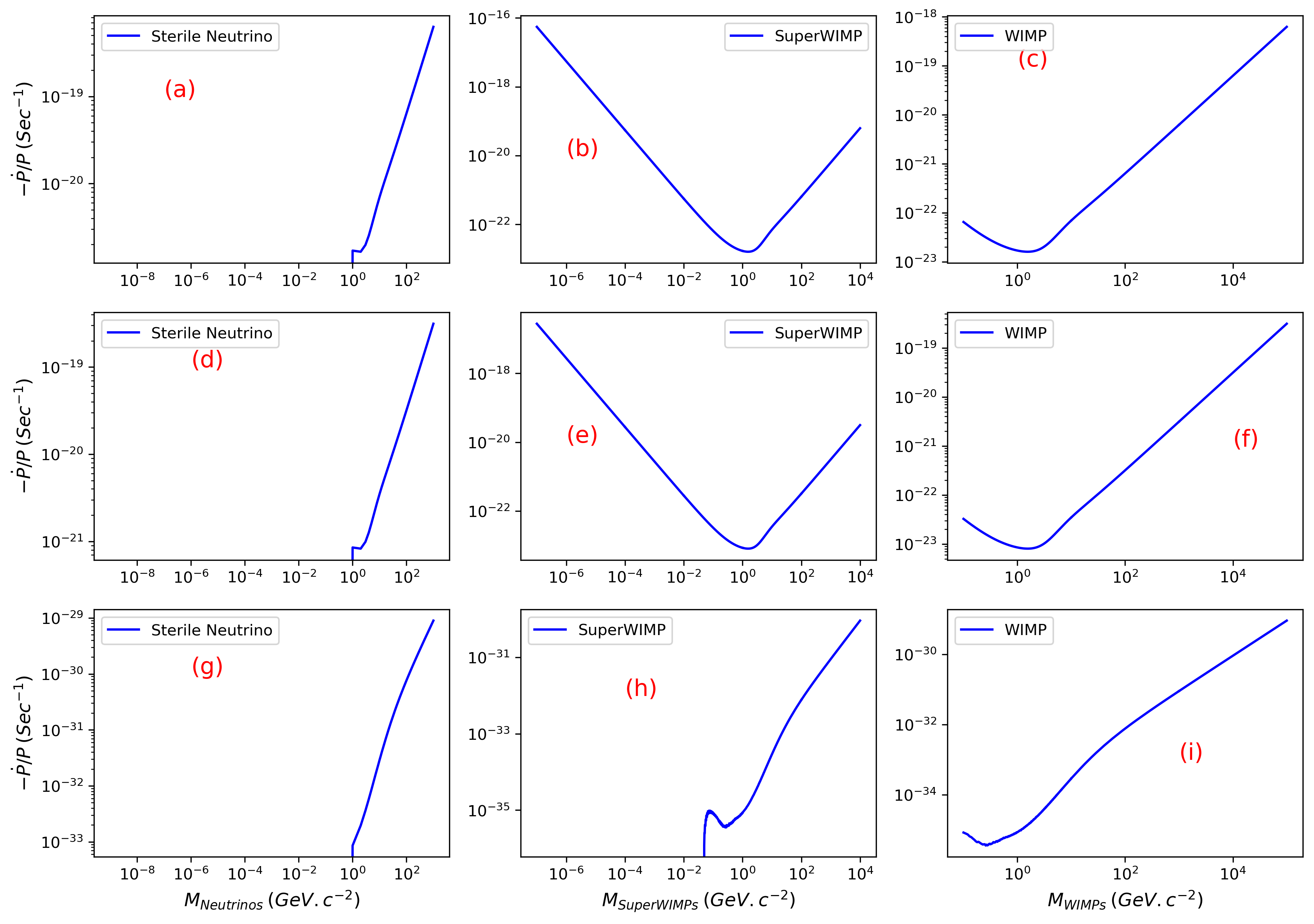}
\caption{The estimated period decay rate for various types of compact binary systems and different DM particle candidates. Sub-plots (a)-(c) are for WD-WD type, sub-plots (d)-(f) for WD-NS type, and sub-plots (g)-(i) are for NS-NS type compact binary systems. In all sub-plots, horizontal axes are the mass range of different DM particle candidates, and vertical axes are the period decay rate of the compact binary systems.}
\label{Fig_CR_All_Systems}
\end{figure*}
\begin{table*}
\caption{Physical parameters of the compact binary systems that are used in this study. Columns include: Type: types of the components of the compact binary systems; Name: name of the systems;  $ M_{p} $ and $ M_{c} $: mass of primary and companion stars; P: orbital period of the systems; d: distance from the galactic center; $\dot{P}^{obs}/P$: measured period change; $\dot{P}^{GW}/P$ and $\dot{P}^{DF}_{NFW}/P$: period change due to the gravitational wave emission and dynamical friction; $\dot{P}_{{\rm Sterile Neutrino}}^{\rm DM}/P$, $\dot{P}_{{\rm SuperWIMP}}^{\rm DM}/P$ and $\dot{P}_{{\rm WIMP}}^{\rm DM}/P$ are estimated period changes of the systems due to the capture of DM particles. All data (except the last three columns) are from Table 1 of the Ref. \cite{PhysRevD.96.063001}.}
\begin{threeparttable}
\resizebox{\textwidth}{!}{%
\begin{tabular}{lllllllllllll} \hline \hline
Type & Name & \shortstack{$m_{p}$ \\ $[M_{\odot}]$} & \shortstack{$m_{c}$ \\ $[M_{\odot}]$} & \shortstack{ $P$ \\ $[days]$} &  \shortstack{d \\ $[kpc]$} & \shortstack{$\dot{P}^{\rm obs}/P$ \\ $ \times (10^{-16})$ \\ $[Sec^{-1}]$ } & \shortstack{$\dot{P}^{\rm GW}/P$ \\ $\times (10^{-16})$ \\ $[Sec^{-1}]$} & \shortstack{$\dot{P}_{{\rm NFW}}^{\rm DF}/P$ \\ $\times (10^{-24})$ \\ $[Sec^{-1}]$} & \shortstack{$\dot{P}_{{\rm Sterile Neutrino}}^{\rm DM}/P$ \\ $[Sec^{-1}]$} & \shortstack{$\dot{P}_{{\rm SuperWIMP}}^{\rm DM}/P$ \\ $[Sec^{-1}]$} & \shortstack{$\dot{P}_{{\rm WIMP}}^{\rm DM}/P$ \\ $[Sec^{-1}]$} \\ \hline
\multirow{5}{*}{\rotatebox{90}{NS-NS}} & J0737-3039 & 1.3381(7) & 1.2489(7) & 0.104 & 1.15(22) & -1.393 & -1.38874 & -1.168 & $10^{-33}-10^{-29}$ & $10^{-36}-10^{-30}$  & $10^{-35}-10^{-29}$  \\ 
& B1534+12 & 1.3330(4) & 1.3455(4) & 0.421 & 0.7 & -0.052905 & -0.037554 & -6.7126  &  &  &   \\ 
& J1756-2251 & 1.312(17) & 1.258(17) & 0.321 & 2.5 & -0.0757 & -0.0793 & -0.009771 &  &  &  \\ 
& J1906+0746 & 1.323(11) & 1.290(11) & 0.166 & 5.4 & -0.3939 & -0.363 & -0.18512  &  &  &  \\ 
& B1913+16            & 1.4398(2) & 1.3886(2) & 0.325 & 9.9            & -0.8533    & -0.8556022    & -0.2828   &  &  &  \\ \hline
\multirow{4}{*}{\rotatebox{90}{NS-WD}} & J0348+0432         & 2.01(4)     & 0.172(3)  & 0.104 & 2.1(2)        & - 0.3038   & -0.2871         & -0.04441  & $10^{-21}-10^{-18}$ & $10^{-23}-10^{-17}$ & $10^{-23}-10^{-18}$ \\ 
& J1012+5307         & 1.64(22)   & 0.16(2)    & 0.60   & 0.836(80)   & -0.289     & -0.0212          & -0.06566  &  &  &  \\ 
& J1141-6545          & 1.27(1)     & 1.02(1)    & 0.20   & 3.7            & -0.2321   & -0.2332          & -0.2071  &  &  &   \\ 
& J1738+0333         & 1.46(6)     & 0.181(7)  & 0.354 & 1.47(10)     & -0.0084680   & -0.009155  & -0.0693  &  &  &  \\  \hline
WD-WD & WDJ0651+2844 & 0.26(4) & 0.50(4) & 0.008 & 1 & -140 & -120 & -0.02025  & $10^{-21}-10^{-18}$ & $10^{-23}-10^{-16}$ & $10^{-23}-10^{-18}$ \\ \hline
\end{tabular}}
\end{threeparttable} \label{table_CR_in_known_binary_systems} 
\end{table*}
\section{Dark matter density profile} \label{Sec: Dark Matter Density Profile}
\begin{figure*}
	\centering
\includegraphics[width=0.8 \columnwidth]{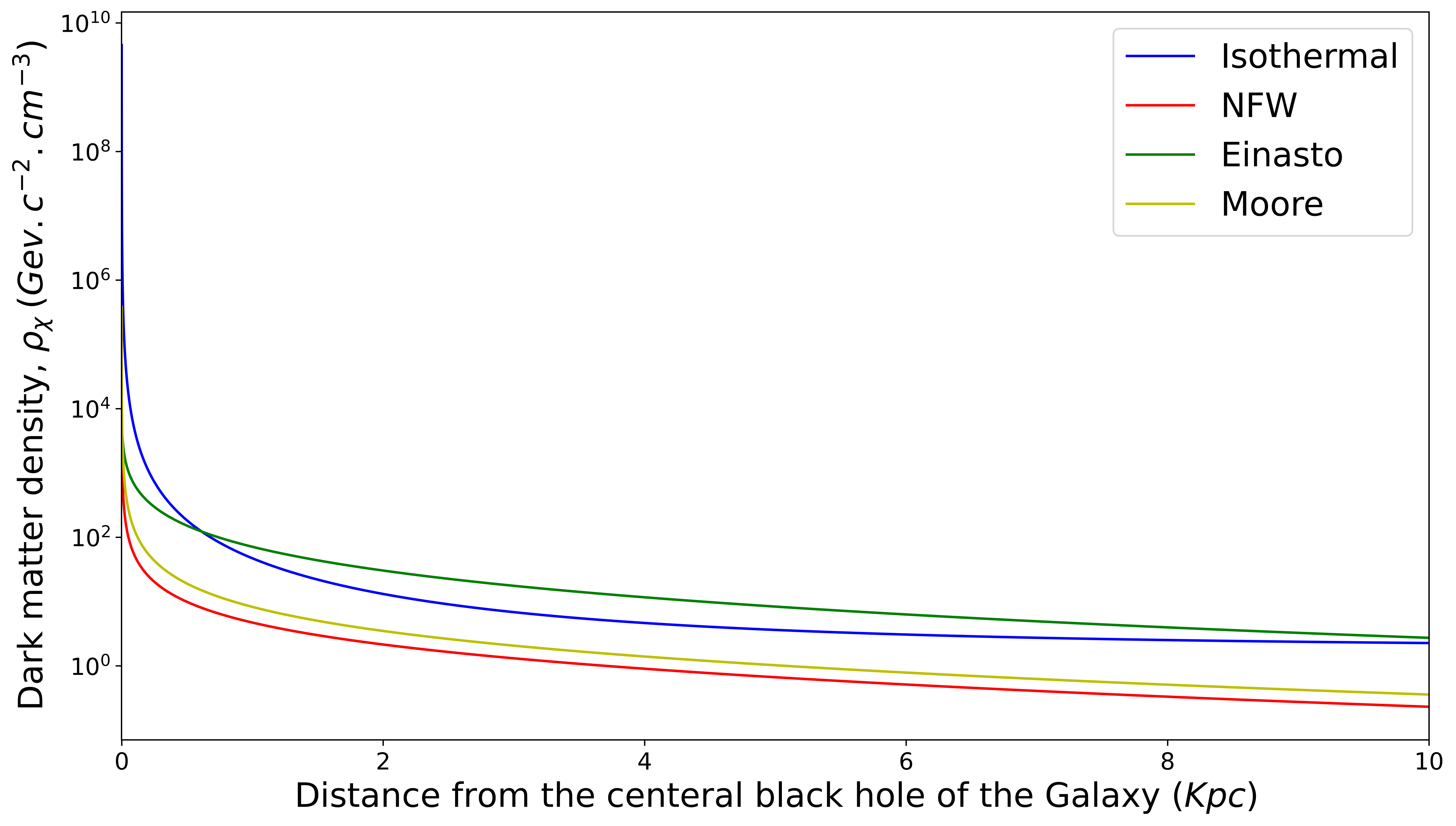}
	\caption{ (color on-line) 2D illustration of the different DM density profiles that are discussed in the text. For distances far from the galactic center (e.g., distances more than 1 kpc), the differences between different DM density profiles are less than one order of magnitude. For this reason, the NFW DM density profile is selected for this study.}
\label{Fig_Dark_Matter_Density_Profile}
\end{figure*}

N-body simulations of galaxies reveals the non-uniform distribution of DM inside galaxies. These results are in agreement with observations (i.e. the rotation curves of galaxies). Navarro, Frenk and White (NFW) DM profile describes the radial distribution of DM density inside galaxies \cite{Merritt2006a,1996ApJ...462..563N}
\begin{equation} 
\rho_{NFW}(r) = \dfrac{\rho_{0}}{\dfrac{r}{r_{s}}(1+\dfrac{r}{r_{s}})^{2}},
\end{equation} \label{Eq: NFW_Profile}
where $ \rho_{0} $ is the local DM density and $ r_{s} $ is the size of the DM halo \cite{Barger2009}. Both $ \rho_{0} $ and $ r_{s} $ vary from a galaxy to a galaxy. In the case of the Milky way galaxy, these parameters are estimated to be about $ \rho_{0} = 0.26 \: (GeV.cm^{-3}) $ and $ r_{s} = 20 \: (kpc) $ \cite{Barger2009} and for the case of M31 galaxy (andromeda galaxy) the estimation for these parameters are about $ \rho_{0} = 0.42 \: (GeV.cm^{-3}) $ and $ r_{s} = 16.4 \pm 1.5 \: (Kpc) $ \cite{Tamm2012}.\\
In addition to the NFW DM density profile, Einasto, Moore, and isothermal DM density profiles can be used to describe the distribution of DM inside galaxies. These profiles have the following expressions \cite{Barger2009}
\begin{equation}
\rho_{\chi, Einasto} = 
\begin{cases}
 & \rho_{0} \: exp \left \{ \frac{-2}{\alpha} \left [ (r^{\alpha} - r^{\alpha}_{s})/r^{\alpha}_{s} \right ) ] \right \} \\
 & \alpha = 0.17, r_{s}=25 \: (kpc),
\end{cases}
\end{equation} \label{Eq: DM_density_profiles}
\begin{equation}
\rho_{\chi, Moore} = 
\begin{cases}
 & \rho_{0} (\frac{r_{\odot}}{r})^{1.16} (\frac{1+r_{\odot}/r_{s}}{1+r/r_{s}})^{1.84} \: \\
 & r_{s}=30 \: (kpc).
\end{cases}
\end{equation} \label{Eq: DM_density_profiles}
\begin{equation}
\rho_{\chi, Isothermal} =
\begin{cases}
&  \rho_{0} \frac{1+(r_{\odot}/r_{s})^{2}}{1+(r/r_{s})^{2}} \: \\
&  r_{s}=5 \: (kpc),
\end{cases}
\end{equation} \label{Eq: DM_density_profiles}
Fig. (\ref{Fig_Dark_Matter_Density_Profile}) depicts the 2D representation of the distribution of DM density according to the NFW, Einasto, Moore, and isothermal DM density profiles and for the central region of our galaxy. From Fig. (\ref{Fig_Dark_Matter_Density_Profile}), it is clear that DM density near the central regions of the galaxies is the highest, and by moving toward the outer regions, its value decreases. Hence, one can conclude that DM affects the physics of all astronomical objects (especially compact binary systems, which is the subject of this study) in the central regions of galaxies more than the objects in the outer regions.\\
For distances far from the central black hole of the galaxy (e.g., farther than 1 kpc), the differences between different DM density profiles are less than one order-of-magnitude. For this reason, we choose the NFW DM density profile throughout the rest of the study. In our calculations, DM density around compact binary systems is considered to be $ \rho_{\chi} = 10^{2} \: (GeV.cm^{-3}) $ which corresponds to the distance of about $ r \simeq 1 \: (kpc) \simeq 3300 \: (ly) $ from the central black hole of the Galaxy. As DM effects on compact binary systems in high DM density environments are the highest, then we selected a distance near the central regions of the Galaxy. By this selection, we studied the highest possible DM effects on compact binary systems.
\section{Results and Discussions} \label{Sec_Results_and_discussion}
We have used Eqs. (\ref{Eq: CR_by_hevier}) and (\ref{Eq_CR_by_binary_systems}) to estimate the accretion of DM particles inside compact binary systems. Stars with masses lower than $1.24 \: M_{\odot}$, and stars with masses higher than $1.24 \: M_{\odot}$ have been considered to be white dwarfs and neutron stars, respectively. The masses of the DM particles have been taken to be in the range $ 10^{-9}-10^{5} \: (Gev.c^{-2}) $ corresponding to the different DM particle candidates that we have used (see horizontal axes of Fig. (\ref{Fig_CR_All_Systems}) for the mass range of different DM particles candidates). \\
From Eq. (\ref{Eq_Keplers_third_law}) we can see that semi-major axis and period of a binary system has the $ a \propto M^{1/3} $ and $ P \propto M^{-1/2} $ proportionality with the total mass of the system. This means by increasing the total mass of the binary system $ M $, the period of the system will be affected more than the semi-major axis. For this reason the $\dot{a}/a$ term can be approximately omitted from the right hand side of the Eq. (\ref{Eq_CR_by_binary_systems}). Then, Eq. (\ref{Eq_CR_by_binary_systems}) can be written in the form
\begin{equation} \label{Eq_CR_by_binary_systems_b}
\frac{\dot{M}}{M} \simeq -2 \frac{\dot{P}}{P},
\end{equation}
which means $\dot{M}/M$ has a linear relation with $\dot{P}/P$. After estimating $\dot{M}/M$ values theoretically (through the procedure that is explained in Sec. (\ref{Sec_CR_By_Compact_Binary_Systems})), we have compared the results with the observed $\dot{P}/P$ values (in the rest of this section we use $\dot{P}^{obs}/P$ notation instead of $\dot{P}/P$ for the observed period changes of compact binary systems). The results of the calculations are presented in Fig. (\ref{Fig_CR_All_Systems}) and table (\ref{table_CR_in_known_binary_systems}). Our main results are:
\begin{itemize}
\item According to table (\ref{table_CR_in_known_binary_systems}), observed values of $\dot{P}/P$  are of the $ 10^{-16}-10^{-19}$ ($ sec^{-1}$) order of magnitude which is almost in the same order of magnitude as the estimated $\dot{P}^{GW}/P$ values. Therefore, we can infer that gravitational wave emission from compact binary systems can be considered as one of the main reasons for the period decay in these systems.
\item In table (\ref{table_CR_in_known_binary_systems}), estimated values of $\dot{P}^{DF}/P$ are of the order of magnitude $ 10^{-24}-10^{-27}$ ($ sec^{-1}$) which are much less than the measured $\dot{P}^{obs}/P$ values. So, we can conclude that dynamical friction that is imposed by DM halo can not be considered as the main reason for the period change in compact binary systems.\\
\item According to the estimated $\dot{P}^{DM}/P$ values in Fig. (\ref{Fig_CR_All_Systems}), period change due to the accretion of DM particles is highly model dependent. Here, we only consider three DM particle models which are Sterile Neutrinos with mass in the range $ 10^{-9} - 10^{3} (GeV.c^{-2}) $, SuperWIMPs with mass in the range $ 10^{-6} - 10^{4} (GeV.c^{-2}) $, and WIMPs with mass in the range $ 10^{-1} - 10^{5} (GeV.c^{-2}) $. Though there are many other DM candidates in the literature (for instance WIMPzilla particles with mass in the range $ 10^{12} - 10^{15} (GeV.c^{-2}) $, \cite{Schumann2019}), for the purposes of this study, using the above mentioned candidates is sufficient. \\

\item From table (\ref{table_CR_in_known_binary_systems}) and Fig. (\ref{Fig_CR_All_Systems}) it is conceivable that binary systems with at least one WD component (i.e. sub-lots (a)-(f) in Fig. (\ref{Fig_CR_All_Systems})) are more affected than binary systems that both components are NS (i.e. sub-plots (g)-(i) in Fig. (\ref{Fig_CR_All_Systems})). This behavior can be understandable from CR Equation too. Using Eq. (\ref{Eq: CR_by_hevier}), numerical estimation gives:
\begin{equation} \label{Eq_CR_Ratio_WD_NS}
\frac{CR_{WD}}{CR_{NS}} \simeq 10^{10} - 10^{18}.
\end{equation}
In Eq. (\ref{Eq_CR_Ratio_WD_NS}) the range of the $CR_{WD}/CR_{NS}$ ratio depends on the DM model. But, and generally speaking, CR by WD stars is many orders of magnitude higher than the CR by NSs.

\item Paying attention to the table (\ref{table_CR_in_known_binary_systems}), we can say that CR of DM particles can not be considered as the main reason for the period change in NS-NS systems. But in binaries with at least one WD, CR of DM particles can be considered as a source of period decay in these systems. This is because, and as discussed above, CR by WDs is about $10^{10} - 10^{18}$ orders of magnitude higher than the CR by NSs.

\item According to Eqs. (\ref{Eq_CR_by_binary_systems_b}) and (\ref{Eq: CR_by_hevier}) period decay of compact binary systems due to the accretion of DM particles is not a monotonic function of DM particles' mass $m_{\chi }$. This behavior can be conceivable from Fig. (\ref{Fig_CR_All_Systems}) too. Variation of other physical parameters (e.g. spin-dependent and spin-independent scattering cross-section, DM density at the location of the binary system, velocity of stars relative to the DM halo, and etc.) can affect the CR amount in compact binary systems. But as the aim of this study is to show that the presence of DM can affect the physical parameters of compact binary systems (and not to study the effects of other physical parameters that affect the CR amount by compact binary systems), then we considered other parameters that affect the CR amounts by compact binary systems to be constant. Investigating the effects of other physical parameters that affect the CR amount by compact binary systems can be a subject of a new and independent study in this respect.
\end{itemize}

Our main result from the above discussions is that the accretion of DM particles into compact binary systems can be considered as one of the main reasons for the observed period decay in WD-WD and WD-NS compact binary systems. But accretion of DM particles into NS-NS binary systems does not affect the orbital parameters of these systems considerably (in comparison to the observed period decay values for NS-NS binary systems).

\section{Data availability}
No new data were generated or analyzed in support of this research.
\section{Acknowledgements}
The authors would like to express their special thanks to Prof. Joakim Edsjö from the University of Stockholm, Sweden, and Prof. Gianfranco Bertone from the University of Amsterdam, Netherlands, and Marco Taoso from the National Institute of Nuclear Physics (INFN) Turin, Italy, for their helpful discussions. We appreciated the comments of the anonymous referee, which resulted in an improved paper. Finally, we acknowledge the use of the Python scientific computing packages NumPy \cite{harris2020array}, SciPy \cite{2020SciPy-NMeth} and Pandas \cite{reback2020pandas}, as well as the graphics environment Matplotlib \cite{Hunter:2007}.

\bibliographystyle{ieeetr}
\bibliography{References_Manuscript}
\end{document}